\documentclass[aps,prl,twocolumn,superscriptaddress]{revtex4}
\usepackage{graphics}
\usepackage{longtable}
\usepackage[epsfig]{graphicx}
\usepackage{epsfig}
\usepackage{amsmath}
\usepackage{float}
\usepackage{placeins}
\usepackage{tikz}         
\makeatletter
\newlength\xvec@height%
\newlength\xvec@depth%
\newlength\xvec@width%
\newcommand{\xvec}[2][]{%
	\ifmmode%
	\settoheight{\xvec@height}{$#2$}%
	\settodepth{\xvec@depth}{$#2$}%
	\settowidth{\xvec@width}{$#2$}%
	\else%
	\settoheight{\xvec@height}{#2}%
	\settodepth{\xvec@depth}{#2}%
	\settowidth{\xvec@width}{#2}%
	\fi%
	\def\xvec@arg{#1}%
	\def\xvec@dd{:}%
	\def\xvec@d{.}%
	\raisebox{.2ex}{\raisebox{\xvec@height}{\rlap{%
				\kern.05em
				\begin{tikzpicture}[scale=1]
				\pgfsetroundcap
				\draw (.05em,0)--(\xvec@width-.05em,0);
				\draw (\xvec@width-.05em,0)--(\xvec@width-.15em, .075em);
				\draw (\xvec@width-.05em,0)--(\xvec@width-.15em,-.075em);
				\ifx\xvec@arg\xvec@d%
				\fill(\xvec@width*.45,.5ex) circle (.5pt);%
				\else\ifx\xvec@arg\xvec@dd%
				\fill(\xvec@width*.30,.5ex) circle (.5pt);%
				\fill(\xvec@width*.65,.5ex) circle (.5pt);%
				\fi\fi%
				\end{tikzpicture}%
	}}}%
	#2%
}
\makeatother

\usepackage{color,soul}

\begin{document}
	
\title{The generalized spin-orbit interaction: a microscopic origin of the \O{}rsted magnetic field}

\author{Sherif Abdulkader Tawfik}
\email{sherif.tawfic@gmail.com}
\affiliation{School of Science, RMIT University, GPO Box 2476, Melbourne, Victoria 3001, Australia}

\begin{abstract}

This work introduces a generalization of the form of the spin-orbit interaction, the generalized spin-orbit interaction (GSOI). It expresses the magnetic field induced by two charged particles moving with a non-zero relative velocity as a field defined at all points in space, and exists in the reference frames of both particles. This is in contrast to spin-orbit interaction theory, in which the generated magnetic field is defined at only one point in space, and exists in the reference frame of one of the two particles. At the macroscopic scale, it is shown that the GSOI theory implies the same form of the \O{}rsted magnetic field produced by a current-carrying wire. However, the theory is incompatible with the microscopic form of the Biot-Savart equation that implies that a charged particle induces a magnetic field by having a non-zero velocity. The implications of the GSOI theory on properties of the \O{}rsted magnetic field in current-carrying atomically thin two-dimensional materials, such as graphene, are discussed. The framework established in this paper aims at re-imagining classical physical concepts in light of an advanced microscopic understanding.
\end{abstract}

\maketitle
	
\section{Introduction}

The Ampere's circuital law (ACL) describes the phenomenon of the creation of the \O{}rsted magnetic field due to the flow of an electric current through a current carrying wire, and is described by the equation $\textbf{J}=\nabla \times \textbf{B}$ \cite{Jackson,Maxwell1}. This law has been verified numerous times and has become deeply entwined in the fabric of fundamental and applied physics. It is an essential ingredient of our present understanding of the connection between the \O{}rsted magnetic field and the electric current. With Maxwell's addition of the displacement field term to the ACL equation, the resulting equation, Maxwell-Ampere equation, in addition to the Maxwell-Faraday equation, form a closed (yet over-determined) set of equations that constitute the classical theory of electromagnetism. ACL was even used for the definition of the \textit{ampere} in terms of the induced magnetic force between two infinitely long, infinitely thin wires \cite{SI}. However, going back to the original equation, ACL, a question that have never been asked in the literature, to the best of my knowledge, is: given that ACL is validated for macroscopic objects, \textit{is it also validated for microscopic objects?}

The first direct application of ACL to a microscopic conducting object was reported by Tetienne \textit{et al.} \cite{NV} two years ago, where the object is a monolayer graphene nanoribbon. The induced magnetic field was measured by the nitrogen vacancy defect in diamond, and the current density was then reconstructed from the magnetic field by inverting the Biot-Savart equation, using the model developed by Roth \textit{et al.} \cite{2DCurrent} for thin slab conductors. Tetienne \textit{et al.} verified the calculated current density by integrating it, which reproduced the value of current that was initially injected into the graphene nanoribbons. However, according to the authors, the system of equations that was used to reconstruct the current density is over-determined, such that solving any two of the four equations yields results that are different from solving by other two of equations. This clearly means that the model does not accurately describe the relationship between the current and \O{}rsted magnetic field in such systems. Importantly, a recent work by Tetienne \textit{et al.} \cite{NV_Problem} directly revealed an inconsistency between the Biot-Savart law and the experimental measurements. This calls for a revision of the fundamental equations describing the induction of the \O{}rsted magnetic field in microscopic conductors.


In fact, the application of ACL to macroscopic wire conductors has not been free from controversy. ACL was shown to suffer a logical deficiency, which is known as the \textit{Ampere tension} \cite{Tension,Tension2,Tension3}: for an electric current flowing through a solid conductor, the application of ACL to a conductor will divide the conductor into longitudinal current elements. Each such element induces the \O{}rsted magnetic field, and therefore applies a repulsive force on the neighboring elements. For any amount of current, this repulsion should lead to longitudinal forces between the current elements, or Ampere tension, that would cause the explosion of the solid conductor. This obviously defies observation. While Graneau \cite{Tension3} reasoned that this tension is the cause of explosion of solid conductors under very high current, and that the size of the smallest possible partitioning of the wire should be that of the crystal lattice unit cell, the existence of an Ampere tension necessitates the existence of lattice strain in the wire during the passage of electric current, no matter how small the current is. This, however, \textit{has never been reported experimentally}. With the large supercurrent densities in superconductors, the Ampere tension would certainly entail the explosion of any superconductor.

The microscopic reduction of ACL is obtained by the Biot-Savart equation. This is derived by taking the current density $\textbf{J}$ as representing the velocity of a single charged particle, $q\textbf{v}$, where $q$ is the charge. For a non-magnetostatic systems, the Jefimenko equations \cite{Jefimenko} are obtained. These two equations were \textit{never verified} for the case of an electron beam in vacuum. This is surprising, given that electron beams are fundamental to many devices. The Ampere tension applies to the electrons flowing in an electron beam: the Jefimenko magnetic field induced by each electron will repel the field induced by the neighboring electrons, which would imply that electrons in an electron beam will deflect sideways, which again defies observation.

This work attempts to establish a relationship between current and the \O{}rsted magnetic field from first principles, based on the spin-orbit interaction (SOI) theory. The SOI is a relativistic effect \cite{Jackson} in which a magnetic field couples to the spin of two charged particles when they move relative to each other. The application of SOI theory to a two-dimensional electron gas by Dresselhaus \cite{Dresselhaus} and Rashba \cite{Rashba} has unraveled the implications of SOI on electronic transport in semiconductors, giving rise to the anomalous Hall effect (AHE) \cite{AHE}, the spin Hall effect (SHE) \cite{SHE1,SHE2} and current-induced spin polarization (CISP) \cite{CISP3}. SOI also gives rise to interesting transport properties in metals, such as the anisotropic magnetoresistance in ferromagnets \cite{AMR} and the SHE in metals \cite{SHE3}. A critical ingredient in these technological advances has been the development of accurate theoretical models for describing the SOI in various systems, such as Rashba's phenomenological model and the incorporation of SOI into density-functional theory (DFT) implementations. Here, I present a generalization of the SOI, the generalized SOI (GSOI), which extends the applicability of the SOI theory beyond the field of spintronics. The GSOI theory can explain the induction of a magnetic field in conducting wires due to the passage of an electric current, and is applicable to the macroscopic as well as the microscopic scales.

\section{The spin-orbit and the spin-other-orbit interaction}  

For two charged particles, $A$ and $B$, with different velocity vectors $\textbf{v}_A$ and $\textbf{v}_B$, the SOI, as derived from the special theory of relativity, is the interaction between the velocity of $A$ and the magnetic field that $A$ experiences in its own rest frame. This induced magnetic field will couple with the spin of $A$, $\mu_A$, according to the spin-orbit coupling energy,
\begin{equation}
H_{SOI}=-\mu_{A} \cdot \textbf{B}_{A}.
\label{eq:H_SOI}
\end{equation}

\noindent The induced magnetic field in the rest frame of particle $i$ ($i=A,B$), $\textbf{B}_{i}$, is given by 
\begin{eqnarray}
\textbf{B}_{SOI,i}&=&-\frac{1}{c^2}(\textbf{v}_i-\textbf{v}_j)\times \textbf{E} \\ \nonumber
&=&-\frac{1}{c^2}(\textbf{v}_i-\textbf{v}_j)\times q_j\frac{\textbf{r}_i-\textbf{r}_j}{\left| \textbf{r}_i-\textbf{r}_j\right| ^3} 
\label{eq:SOI}
\end{eqnarray}
\noindent where $\textbf{v}_i$ is the velocity of particle $i$ and $c$ is the speed of light. For illustration, the schematic in Fig. \ref{fig1}(a) shows the direction of the field induced in the reference frame of an electron moving next to an atom, as well as the field induced in the reference frame of the atom due to the motion of the electron.
	
The magnitudes and directions of $\textbf{B}_{SOI,A}$ and  $\textbf{B}_{SOI,B}$ depend on the charges of the two particles, $q_A$ and $q_B$. If $q_A=q_B$, then $\textbf{B}_{SOI,A}=\textbf{B}_{SOI,B}$, such as for the case of two electrons moving relative to each other. For an electron moving with respect to an ion and experiencing its effective nuclear charge, the two fields are opposite in direction, and the magnitude of the field acting at the position of the electron will be larger than that acting at the position of the ion.

The field $\textbf{B}_{SOI,A}$ has has no source, and there are no field lines except the field vectors defined at the positions of particle $A$ and $B$. The theory of the SOI only accounts for the effect of the transformed field at the position of the particles, but does not aim at establishing a vector field in the reference frames of the particles.

The SOI coupling hamiltonian in Eq. \ref{eq:H_SOI} can be derived from the solution of the Dirac equation up to $1/c^2$ order. The Dirac hamiltonian was then extended by Breit \cite{Breit} to account for the effect of retardation on two Dirac particles. Of interest to the present work is the spin-other-orbit interaction (SOOI) in the Breit hamiltonian $H_{SOOI}$ for two particles $A$ and $B$, which is given by \cite{Breit2}

\begin{equation}
-\frac{e^2}{m^2c^2} \frac{1}{4\pi\epsilon_0}
\frac{ \pmb{\mu}_B\cdot (\textbf{r}_A-\textbf{r}_B)\times \textbf{p}_A+  \pmb{\mu}_A\cdot (\textbf{r}_B-\textbf{r}_A)\times \textbf{p}_B}{\left|\textbf{r}_A-\textbf{r}_B\right|^3}
\label{eq:Breit}
\end{equation}

\noindent where $\textbf{p}_i$, is the particle's momentum vector, $i=A,B$, $\textbf{r}_i$ is its position vector, and $\pmb{\mu}_i$ is its spin operator vector. This terms can be interpreted as the energy required to align the spin operator vector of each particle with the magnetic field induced by the other particle (that is why it is called spin-other-orbit interaction). 

This property makes $H_{SOOI}$ fundamentally different from $H_{SOI}$ in Eqs. \ref{eq:H_SOI} and \ref{eq:SOI}. In SOI, the induced magnetic field at the reference frame of particle $A$ couples with the spin of particle $A$, $\pmb{\mu}_A$ but this field is not observable in the reference frame of particle $B$. In the case of $H_{SOOI}$ in Eq. \ref{eq:Breit}, the magnetic field induced in the reference frame of particle $A$ couples with the spin of particle $B$, $\pmb{\mu}_B$. This means that the Breit interaction theory implicitly makes the induced SOI magnetic field observable in the reference frames of \textit{both} particles.

Based on the above, I propose a generalization for the form of the magnetic field induced by the SOI, the generalized spin-orbit interaction (GSOI) theory.

\section{The generalized spin-orbit interaction theory}

In the GSOI theory, the magnetic field vector induced by the motion of a charged particle relative to another charged particle, $\textbf{B}_{SOI,ij}$ (Eq. \ref{eq:SOI}), is a magnetic field that exists in the reference frame of each of the two particles. That is, the field $\textbf{B}_{SOI,ij}$ in the reference frame of one particle, $A$, is observable in the reference frame of the other particle, $B$. The form of the field $\textbf{B}^0_{GSOI,ij}$ is mathematically similar to that of $\textbf{B}_{SOI}$:

\begin{eqnarray}
\label{eq:GSOI}
\textbf{B}_{GSOI,ij}(\textbf{r})&=&  \textbf{B}_{SOI,ij}  B_{d,ij}B_{r}(\left| \textbf{r}_i-\textbf{r}\right|) \quad, \\
B_{d,ij}&=&e^{-k\left| \textbf{r}_i-\textbf{r}_j\right|} \quad,  \\
B_{\rm{rad}}(\left| \textbf{r}_i-\textbf{r}\right|) &=& \frac{ \kappa }{\left| r\right| +\kappa}
\end{eqnarray}

\noindent where $q_j$ is the charge of particle $j$, $\textbf{B}_{SOI,ij} $ is given by Eq. \ref{eq:SOI}, $B_{d,ij}$ is the exponential decay of the electric field inside a conductor due to charge screening, $B_{\rm{rad}}(r)$ represents the radiation of the SOI magnetic field line, and $\kappa$ is a decay constant in units of $1/\AA$.

The form of $\textbf{B}_{GSOI,ij}(\textbf{r})$ satisfies the following two conditions:

\begin{enumerate}
	\item At the positions $\textbf{r}_A$ and $\textbf{r}_B$ of two charged particles $A$ and $B$ moving with non-zero relative velocity, $\textbf{B}_{GSOI}(\textbf{r})_i$ takes the form of Eq. \ref{eq:SOI} by substituting $\textbf{r}=\textbf{r}_i$. That is, the GSOI field reduces to the SOI field at the positions of the two particles inducing the field.
	\item For positions in space other than $\textbf{r}_A$ and $\textbf{r}_B$,  $\textbf{B}_{GSOI}(\textbf{r})_i \sim 1/r$, where $r=\left|\textbf{r} \right|$.
\end{enumerate}

The GSOI field $\textbf{B}_{GSOI,ij}(\textbf{r})$ complements the theory of SOI by defining a magnetic radiation field. The radiation \textit{source} is the particle that is experiencing a point $\textbf{B}_{SOI,ij}$ magnetic field.

\begin{figure}[h]
	\includegraphics[width=80mm]{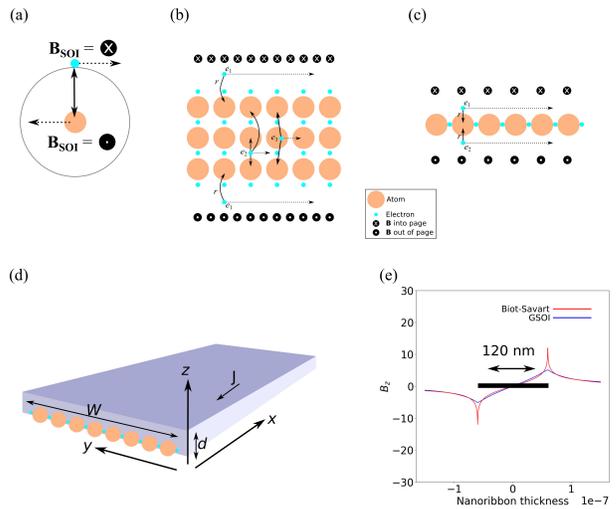}
	\caption{\label{fig1} (a) A schematic illustration of the direction of the magnetic field induced by the standard SOI. (b-d) A schematic illustration of the GSOI theory. (b) The GSOI theory as applied to a either current-carrying conducting wire or a conducting slab. The motion of the free electrons relative to the stationary ions, as well as the relative direction of the two electrons with respect to each other (as indicated by the arrow on the line connecting them) generates the GSOI magnetic field according to Eq. \ref{eq:GSOI}. In the top layer, the GSOI field lines are pointing into the page, whereas in the bottom layer, the lines are pointing out of the page. The GSOI field is generated by the interaction between the conduction electron and the surface electrons closest to it. The contribution of the electrons $e_1$, $e_2$ and $e_3$ to the GSOI field are different, as elaborated in the text. (c) The GSOI field induced by an electric current flowing through an atomically thin conductor, such as an atomic wire or a two-dimensional atomic layer. The cancellation between the GSOI fields on either sides of the conductor leads to the drop of magnetic induction in these conductors. (d) A schematic of a conducting slab, showing the direction of the injected current density and the induced GSOI field. (e) A plot of the $\textbf{B}_{GSOI}$ field versus the $\textbf{B}_{BS}$ field for a thin conducting strip of width 120 nm. $B_z$ is in units of $\mu$T, and the strip width in meters.}
\end{figure}

Equation \ref{eq:GSOI} is illustrated by the schematic diagram in Figs. \ref{fig1}(b), showing electrons moving due to the application of an electric potential along the conductor's axis. The motion of the various electrons, labeled as $e_1$, $e_2$ and $e_3$, generates the GSOI field. The $e_1$ electron moves in the charge cloud surrounding the surface of the conductor, $e_2$ moves right below the surface while $e_3$ moves close to the center of the current-carrying wire. The $e_1$ electron experiences an interaction with the ions on the surface of the current-carrying wire, thus generating the GSOI field. The $e_3$ electron, on the other hand, does not contribute to the GSOI field because the generated field due to interactions with ions \textit{above} it cancels out the GSOI field generated due to its interaction with ions \textit{below} it. This cancellation becomes weaker as we approach the surface of the conductor, such as the case of $e_2$. Therefore the motion of the $e_2$ electron generates a non-negligible GSOI field. Thus, the closer an electron is to the surface of the conductor, the larger is the $\textbf{B}_{GSOI}$ field it creates; the closer it is to the center, the weaker is the $\textbf{B}_{GSOI}$ field it creates.

\section{Application to nanomaterials}

For the case reported in Ref. \cite{NV_Problem}, in which a conducting strip had a width of 10 $\mu m$ and thickness of 100 nm, the application of the Biot-Savart law deviated from the experimental measurement. For a conducting strip axis along the $x$-axis and with the $z$-axis normal to its plane, the experimentally measured  $B_{z}$ field at the strip edge is smaller than the  $B_{BS,z}$ calculated by the Biot-Savart law. 

In an attempt to use the GSOI theory to explain the form of the $B_{z}$ field for the conducting strip, Fig. \ref{fig1}(e) compares the $B_{GSOI,z}$ field against the Biot-Savart field  given by $B_{BS,z}(x)=\frac{\mu_0 I}{4\pi w} {\rm ln}\frac{\left| x+w/2\right| }{\left| x-w/2\right| }$, where $I$ is the current flowing along the $y$-axis, and $w$ is the width. The $B_{GSOI,z}$ field at the strip edge is smaller than the $B_{BS,z}$, which is similar to the experimental measurement in Ref. \cite{NV_Problem}. While the plot in Fig. \ref{fig1}(e) is for a strip that is 120 nm, which is much smaller than the strip in Ref. \cite{NV_Problem} that is 10 $\mu$m wide, the behavior of $B_{GSOI,z}$ and $B_{BS,z}$ at the strip edge will be similar because of the asymptotic behavior of $B_{BS,z}$ at the edge.

For the case of graphene, Fig. \ref{fig1}(e) displays the slab structure of a graphene nanoribbon. In a graphene nanoribbon that is suspended in vacuum, there are two conducting systems:
\begin{enumerate}
	\item The two infinitely-thin conducting slabs, which are the $\pi$ bond networks above and below the hexagonal carbon atoms and both are two-dimensional: These two networks are filled by electrons occupying the $p_z$ orbital of the carbon atoms, and therefore a $p_z$ electron is equally likely to occupy either $\pi$ bond network, the top or the bottom. Only the $p_z$ orbitals of graphene are the ones that contribute to the calculation of the SOI in graphene \cite{RashbaGraphene}. Therefore, only the $\pi$ bond networks will contribute to the GSOI field generation.
	\item The nanoribbon edge which is one-dimensional: as discussed above, the edge supports a persisting edge state that sustains the nanoribbon conduction current.
\end{enumerate}

\noindent For a graphene sheet to which the $z$-axis is normal and through which a current flows along the $+x$ direction, the two $\pi$-bond layers each induce a $\textbf{B}_{GSOI,t/b}$ (where $t$ stands for top, $b$ for bottom) that is strictly in the direction of the $+y$-axis. The superposition of the top and bottom $\textbf{B}_{GSOI}$ results in the total induced $\textbf{B}_{GSOI,y}$ component of the field that is less than the $\textbf{B}_{GSOI,y}$ component induced by each $\pi$-bond network alone. The $\textbf{B}_{GSOI,z}$ component of the field, however, is enhanced by the superposition of the GSOI field across the width of the nanoribbon, as is displayed in Fig. \ref{fig1}(h). $\textbf{B}_{GSOI,z}$ is also enhanced by the larger linear current density flowing through the edge of the nanoribbon.

\section{A microscopic origin of Ampere's circuital law}
\label{ACL}

A discussion of a possible microscopic origin for the \O{}rsted magnetic field actually took place in the years following Ampere's publication of the Ampere force law (AFL) \cite{Assis}. AFL proposes an empirical action-at-a-distance formula between the electric current passing through two wires, the distance between the wires and the force induced by each wire on the other. In spite of its simplicity and experimental validation, its interpretation was problematic ever since it was published. While Ampere believed that his equation describes a fundamental law of the nature of the electric current, which he generalized to state that magnetism in magnetic materials arises due to the presence of microscopic currents, Biot and Savart proposed an alternative interpretation, the wire magnetization hypothesis (WMH) \cite{BiotSavart1,BiotSavart2,BiotSavart3}. According to the WMH, the magnetic field in the current-carrying conductor arises because the wire itself becomes magnetized. Thus, the magnetic field does not arise from a moving charge, but instead arises from the lining up of microscopic magnetic particles around the circumference of the wire. This interpretation was criticized by Ampere \cite{Assis}, and was not endorsed by the physics community. If that interpretation was true, however, the idea that a moving charge induces a magnetic field would have to be dropped, since that induction is only a result of the response of the material in the current-carrying conductor to the passing electric current. For the originators of WMH, this conclusion would actually contradict with the application of their own equation (the Biot-Savart equation) to the case of a moving charged point particle.

The WMH that is solely based on the alignment of microscopic magnetic dipoles through the current-carrying conductor is in fact problematic because the resulting force between two current-carrying conductors would scale as $r^{-4}$, which is the force between two dipoles separated by distance $r$, whereas the observed magnetic force between two current-carrying conductors scales as $r^{-2}$.

The GSOI theory proposed in this work derives the \O{}rsted magnetic field from a first principles theory. GSOI therefore re-introduces the WMH in terms of modern theoretical foundations. The GSOI theory proposes that the \O{}rsted magnetic field is a \textit{material}, rather than a \textit{field} property; the field only emerges when an electric current passes through a conducting medium, not solely due to the motion of charge. 

Would the GSOI theory affect the theory of field-particle interaction? This is not the case, because the field-matter interaction term, $q\textbf{p}\cdot \textbf{A}$, where $q$ is the particle's charge, $\textbf{p}$ is the particle momentum and $\textbf{A}$ is the vector potential, is fully derivable by constructing the wave-particle Lagrangian based on three ingredients, the Maxwell-Faraday equation, the relationship $\textbf{B}=\nabla \times \textbf{A}$ and the Lorentz force equation. None of these ingredients depend on the Maxwell-Ampere equation.

\section{Conclusion } 

The generalized spin-orbit interaction (GSOI) theory is proposed. It expresses the magnetic field induced by two charged particles moving with a non-zero relative velocity as a field defined at all points in space. This is in contrast to spin-orbit interaction theory, in which the generated magnetic field is defined at only one point in space, and is observable in the reference frame of only one of the two particles. When applied to a current-carrying wire, the GSOI theory can reproduce the Biot-Savart form of the \O{}rsted magnetic field. Hence, GSOI can potentially replace the Maxwellian conception of \O{}rsted magnetic induction, in which a single moving point charge induces a magnetic field.

\end{document}